# Energy spectrum and optical absorption of superlattices under strong band inversion conditions.

B.Laikhtman[1], S.Suchalkin[2], G. Belenky[2] and M.Ermolaev[2]

[1]Racah Institute of Physics, Hebrew University, Jerusalem 91904, Israel

[1]State University of New York at Stony Brook, Stony Brook, New York, 11794-2350, USA

Abstract

We show that in superlattices with a strong band inversion no hybridization gap exists. There are two points where the bands are crossing and the spectrum has a shape of Dirac cones. Due to the absence of the hybridization gap the optical absorption does not have a threshold and goes to zero gradually with the decrease of optical energy.

Significant interest is arising in the material development and studies of the energy spectra of semiconductors with inverted bands where the top of Γ8 valence band has energy higher than this of the bottom of Γ6 conduction band [1,2]. Structures with inverted bands can be designed [3] and grown using InAs/GaSb superlattices (SLs) [4,5] and coupled QWs [6,7] with various layer thicknesses. In the latter case the transport character is two dimensional. Three-dimensional bulk and quasi-bulk materials with gapless and inverted bands were demonstrated in HgCdTe alloys [8] and HgTe/HgCdTe SLs with different Cd compositions [9].

Recent success in the development of metamorphic heterostructures based on InAs/InSb system [10,11] opens new opportunities for band engineering of narrow gap semiconductors. It was shown experimentally [12] that in the case of strain compensated InAsSb./InSb type II SLs with ultrathin layers variation of only SLs period allows one to produce three dimensional gapless material and SLs with inverted bands [13]. In this case the band gap transformation is achieved as a result of variation of strain and carrier confinement without changes of the material composition. InAsSb/InSb SLs can be a potential competitor of HgCdTe crystals for very long wave length IR photodetectors and a new platform for realization of nontrivial topological phases and strongly correlated electronic states [14, 15].

In this paper we present the results of theoretical studies of the InAsSb./InSb SL energy spectrum under condition of strong conduction and valence band inversion. Contrary to 2D systems where any conduction and valence band overlap leads to formation of hybridization gap and gapless state exists only at very specific values of parameters when the band gap is zero [16], in SLs there exist two contact points of the upper and low band with Dirac conical spectrum at any band overlap. Possibility of the contact points was pointed out by Altarelli many years ago [17]. The basic reason for the band crossing is a strong anisotropy of the SL spectrum. The overlap of in-plane conduction and valence band spectra is maximal at zero wave vector in the growth direction $q$ and decreases with growth of $q$. If the maximal overlap is smaller than the sum of the upper and lower band width then growing $q$ inevitably passes the value where the overlap of the in-plane spectra is zero, Fig.2a. This value corresponds to contact points.

It makes sense to note that the hybridization gap in the density of states can disappear also in coupled quantum wells.[18] There in case of strong band overlap anisotropy of the valence band becomes important. Due to the anisotropy the hybridization gap at different regions of the Brillouin zone is formed at different energies and the energy gaps may not overlap. But in this case the "local" gap exists in every point of the Brilloin zone and the optical absorption edge is preserved due to vertical character of the optical transitions.

To demonstrate absence of absorption threshold in SLs we make use here a two-band model that makes the phenomenon very transparent. In spite of simplicity of the model comparison of its results with an experiment and a very accurate calculation[19] shows that its inaccuracy is no more than 20% - 30%.

Simple analytic version of *kp* method for heterostructures was developed in Ref.20. According to this method the Hamiltonian of electron and heavy hole band in a SL is

$$H = \begin{pmatrix} \varepsilon_e(q) & \dfrac{\hbar k_+ P}{\sqrt{2m_0}} \\ \dfrac{\hbar k_- P}{\sqrt{2m_0}} & \varepsilon_{hh}(q) \end{pmatrix} \qquad (1)$$

where $\varepsilon_e(q)$ and $\varepsilon_{hh}(q)$ are spectra of the first miniband of conduction and heavy hole valence band in the SL, $q$ is the quasi-wave vector in the growth direction $z$, $k_\pm = k_x \pm i k_y$ and $P$ is the momentum matrix element between the conduction and valence band averaged over the SL period. Hamiltonian (1) leads to the spectrum

$$E_\pm = \frac{\varepsilon_e(q) + \varepsilon_{hh}(q)}{2} \pm \sqrt{\frac{[\varepsilon_e(q) - \varepsilon_{hh}(q)]^2}{4} + \frac{\hbar^2 P^2 k_\parallel^2}{2 m_0^2}} \qquad (2)$$

where $k_\parallel^2 = k_x^2 + k_y^2$. An important point here is that compared to type II coupled quantum wells, the band overlap $\varepsilon_e(q) - \varepsilon_{hh}(q)$ depends on the wave vector $q$ in the growth direction. Overlap of the bands means that $\varepsilon_g = \varepsilon_e(0) - \varepsilon_{hh}(0) < 0$. At the edge of the Brillouin zone $\varepsilon_e(\pi/d) = \varepsilon_e(0) + \Delta_e$ and $\varepsilon_{hh}(\pi/d) = \varepsilon_{hh}(0) - \Delta_{hh}$ where $\Delta_e$ and $\Delta_{hh}$ are widths of electron and heavy hole band and $d$ is the SL period. If $|\varepsilon_g| < \Delta_e + \Delta_{hh}$ than there exists $q = \pm q_c$ where $\varepsilon_e(q_c) - \varepsilon_{hh}(q_c) = 0$. At this value and $k_\parallel = 0$ the hybridization gap disappears. Both $\varepsilon_e(q)$ and $\varepsilon_{hh}(q)$ depend on the material and width of SL layers. Therefore value of $q_c$ depends on the design of the SL.

If the band energy separation or overlap is small it is enough to consider the region $q \ll \pi/d$ where the electron and heavy hole spectra are parabolic,

$$\varepsilon_e(q) = \varepsilon_e(0) + \frac{\hbar^2 q^2}{2 m_{e\perp}}, \qquad \varepsilon_{hh}(q) = \varepsilon_{hh}(0) - \frac{\hbar^2 q^2}{2 m_{hh\perp}} \qquad (3)$$

and Eq.(2) turns into

$$E_\pm = \frac{\varepsilon_e(0) + \varepsilon_{hh}(0)}{2} + \frac{\hbar^2 q^2}{2}\left(\frac{1}{m_{e\perp}} - \frac{1}{m_{hh\perp}}\right) \pm \sqrt{\frac{1}{4}\left(\varepsilon_g + \frac{\hbar^2 q^2}{2 m_{eh\perp}}\right)^2 + |\varepsilon_g|\frac{\hbar^2 k_\parallel^2}{2 m_\parallel}} \qquad (4)$$

where $\varepsilon_g$ is the band gap at $q=0$ and $m_{eh\perp} = m_{e\perp} m_{hh\perp} / (m_{e\perp} + m_{hh\perp})$ is the reduced electron – hole mass. Near the origin of $\mathbf{k}$ space when $q$ and $k_\parallel$ are so small that $\hbar^2 q^2 / 2m_{eh}, \hbar^2 P^2 k_\parallel^2 / 2m_0^2 \ll |\varepsilon_g|$ the spectrum of both upper and low band is quadratic with in-plane effective mass $m_\parallel = |\varepsilon_g| m_0 / P^2$. If the bands don't overlap, $\varepsilon_g > 0$ and the spectrum is parabolic. If the bands overlap $\varepsilon_g < 0$ and the spectrum of each of the band is hyperbolic paraboloid where the origin is the saddle point of the spectra. The double value of the square root in Eq.(4) gives the gap in the spectrum. At $\varepsilon_g > 0$ the gap is positive at any value of $q$ and $k_\parallel$. In the other case, $\varepsilon_g < 0$ there are two band crossing points where the gap disappears: $k_\parallel = 0$, $q = \pm q_c$, $q_c = \sqrt{2 m_{eh} |\varepsilon_g|} / \hbar$. Near the crossing points the spectrum is conical. Typical plot of the spectrum with overlapping bands is shown in Fig.1. In Fig.2 spectrum at different cross sections of surface $E(k_\parallel, q)$ is shown.

The difference between optical absorption in cases of overlapped and non-overlapped minibands and the absence of the threshold in the former case can be seen very clearly when the Fermi level is in between them. In this case optical absorption coefficient is defined by the expression

$$\alpha(\hbar\omega) = \frac{4\pi^2 e^2}{m_0^2 c\omega\sqrt{\kappa}} \sum_s \int_{-\pi/d}^{\pi/d} \frac{dq}{2\pi} \int |M_s|^2 \delta(E_+ - E_- - \hbar\omega) \frac{d^2 k_\parallel}{(2\pi)^2} \qquad (5)$$

where $\kappa$ is the optical dielectric constant averaged over the SL period and the summation is carried out over all spin states. $M_s$ is the matrix element of $\mathbf{ep}$ ($\mathbf{e}$ is the photon polarization vector and $\mathbf{p}$ is the electron momentum) between initial and final wave functions

$$\Psi_{k_\parallel qs-} = \frac{1}{\sqrt{S}} e^{i\mathbf{r}_\parallel \mathbf{k}_\parallel} \left[ \psi_{hh-}(\mathbf{k}_\parallel, q) \xi_{hh,q}(z) u_{hh,s}(\mathbf{r}) + \psi_{e-}(\mathbf{k}_\parallel, q) \xi_{e,q}(z) u_{e,s}(\mathbf{r}) \right]$$
$$\Psi_{k_\parallel qs+} = \frac{1}{\sqrt{S}} e^{i\mathbf{r}_\parallel \mathbf{k}_\parallel} \left[ \psi_{hh+}(\mathbf{k}_\parallel, q) \xi_{hh,q}(z) u_{hh,s}(\mathbf{r}) + \psi_{e+}(\mathbf{k}_\parallel, q) \xi_{e,q}(z) u_{e,s}(\mathbf{r}) \right] \qquad (6)$$

Here $u_{e,s}(\mathbf{r})$ and $u_{hh,s}(\mathbf{r})$ are Bloch amplitudes of the conduction and heavy hole valence band (different in different layers), $\xi_{e,q}(z)$ and $\xi_{hh,q}(z)$ are SL wave functions and $\psi_{e\pm}(\mathbf{k}_\parallel, q) = |\psi_{e\pm}(\mathbf{k}_\parallel, q)| e^{i\varphi_{e\pm}(\mathbf{k}_\parallel)}$ and $\psi_{hh\pm}(\mathbf{k}_\parallel, q) = |\psi_{hh\pm}(\mathbf{k}_\parallel, q)| e^{i\varphi_{hh\pm}(\mathbf{k}_\parallel)}$ are electron and heavy hole components of eigenfunctions of Hamiltonian (1). Phases of these components are connected by the relations $\varphi_{e+}(\mathbf{k}_\parallel) - \varphi_{hh+}(\mathbf{k}_\parallel) = \varphi_{k_\parallel}$ and $\varphi_{e-}(\mathbf{k}_\parallel) - \varphi_{hh-}(\mathbf{k}_\parallel) = \varphi_{k_\parallel} \pm \pi$ where $\tan \varphi_{k_\parallel} = k_y / k_x$ and the moduli are

$$|\psi_{e\pm}(\mathbf{k}_\parallel, q)|^2 = |\psi_{hh\mp}(\mathbf{k}_\parallel, q)|^2 = \frac{1}{2} \pm \frac{\varepsilon_e(q) - \varepsilon_{hh}(q)}{2\sqrt{[\varepsilon_e(q) - \varepsilon_{hh}(q)]^2 + 2\hbar^2 P^2 k_\parallel^2 / m_0^2}} \qquad (7)$$

Integration with respect to $\mathbf{k}_\parallel$ in Eq.(5) can be carried out analytically and the result is

$$\alpha(\hbar\omega) = \frac{e^2}{c\hbar} \frac{\sin^2 \vartheta}{(2\hbar\omega)^2 \sqrt{\kappa}} \int_{-\pi/d}^{\pi/d} \left[ (\hbar\omega)^2 + [\varepsilon_e(q) - \varepsilon_{hh}(q)]^2 \right] \theta(\hbar\omega - |\varepsilon_e(q) - \varepsilon_{hh}(q)|) dq \qquad (8)$$

where $\vartheta$ is the angle between the photon polarization vector and $z$ axis.

At small optical energies wave vectors $q$ and $k_\parallel$ are small and it is possible to make use of quadratic approximation for the spectrum, Eq.(3). Then Eq.(8) is reduced to

$$\alpha(\hbar\omega) = \alpha_0 f_I(\hbar\omega/\varepsilon_g) \qquad \varepsilon_g > 0, \qquad (9a)$$

$$\alpha(\hbar\omega) = \alpha_0 f_{II}(\hbar\omega/|\varepsilon_g|) \qquad \varepsilon_g < 0, \qquad (9b)$$

where

$$\alpha_0 = \frac{e^2}{c\hbar} \frac{\sqrt{2m_{eh}|\varepsilon_g|}\sin^2\vartheta}{2\hbar\sqrt{\kappa}} \qquad (10)$$

and

$$f_I(x) = \theta(x-1)\frac{\sqrt{x-1}}{x^2}\left[x^2+1+\frac{2}{3}(x-1)+\frac{1}{5}(x-1)^2\right]$$

$$f_{II}(x) = \frac{\sqrt{x+1}}{x^2}\left[x^2+1-\frac{2}{3}(x+1)+\frac{1}{5}(x+1)^2\right]-\theta(1-x)\frac{\sqrt{1-x}}{x^2}\left[x^2+1-\frac{2}{3}(1-x)+\frac{1}{5}(1-x)^2\right] \qquad (11)$$

Plot of functions $f_I$ and $f_{II}$ is presented in Fig.3. If bands don't overlap there is a threshold of absorption and near the threshold $\alpha(\hbar\omega) \propto \sqrt{\hbar\omega-\varepsilon_g}$. If bands overlap there is no threshold and $\alpha(\hbar\omega) \propto \hbar\omega$ at small $\hbar\omega$. The origin of a singularity of $f_{II}$ at $x=1$, that is a singularity of $\alpha(\hbar\omega)$ at $\hbar\omega = |\varepsilon_g|$ is the following. At $\hbar\omega < |\varepsilon_g|$ optical transitions are possible at small values of $q$ where bands overlap. At larger values of $\hbar\omega$ this possibility disappears.

**Summary**

It was shown that in superlattices with strong band inversion no hybridization gap exists. At zero in-plane wave vector there are two crossing (contact) points of the upper and lower bands where the carrier spectrum has a form of Dirac cones. Position of the crossing points in the momentum space depends of the design of the SL. Due to the absence of the hybridization gap the optical absorption does not have a threshold and goes to zero gradually with the decrease of optical energy.

**Acknowledgement**

We appreciated helpful discussions with S. Svensson and support of National Science Foundation grant no. DMR-1809708 and of U.S. Army Research Office grant no. W911TA-16-2-0053.

# Figure captions

Fig.1. Typical spectrum of overlapping first SL minibands in conduction and heavy hole valence band calculated according to two band model, Eq.(4).

Fig.2. Energy dependence on $q$ and $k_\parallel$ at different cross sections of surface $E(k_\parallel, q)$ calculated for InSb/InAs0.6Sb SL with the ratio of layer thicknesses 1:1 and the period of ~ 6 nm using 8 band ***kp*** method. (a) $E(0,q)$, (b) $E(k_\parallel, 0)$, (c) $E(k_\parallel, q_c)$, (d) $E(k_\parallel, q_d)$.

Fig.3. Plot of functions $f_I$ and $f_{II}$ that control the energy dependence of the absorption coefficients, Eq.(9). $f_I(\hbar\omega/\varepsilon_g) = 0$ if $\hbar\omega < \varepsilon_g$ and $f_I(\hbar\omega/\varepsilon_g) = 2\sqrt{\hbar\omega/\varepsilon_g - 1}$ near the threshold. $f_{II}$ does not have a threshold and $f_{II}(\hbar\omega/|\varepsilon_g|) = 4\hbar\omega/3|\varepsilon_g|$ at small $\hbar\omega/|\varepsilon_g|$.

Figure 1

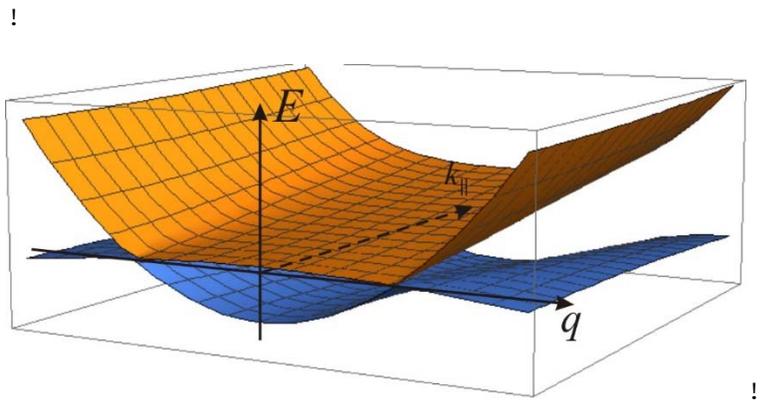

Figure 3

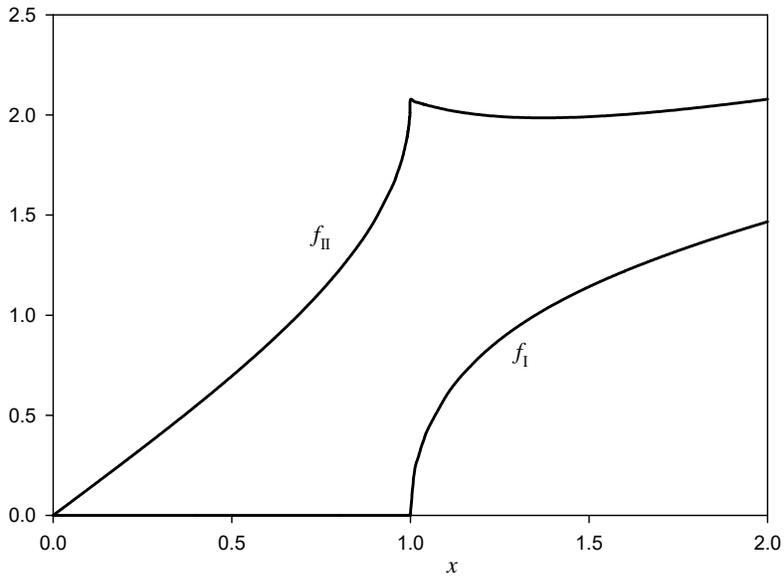

Figure 2

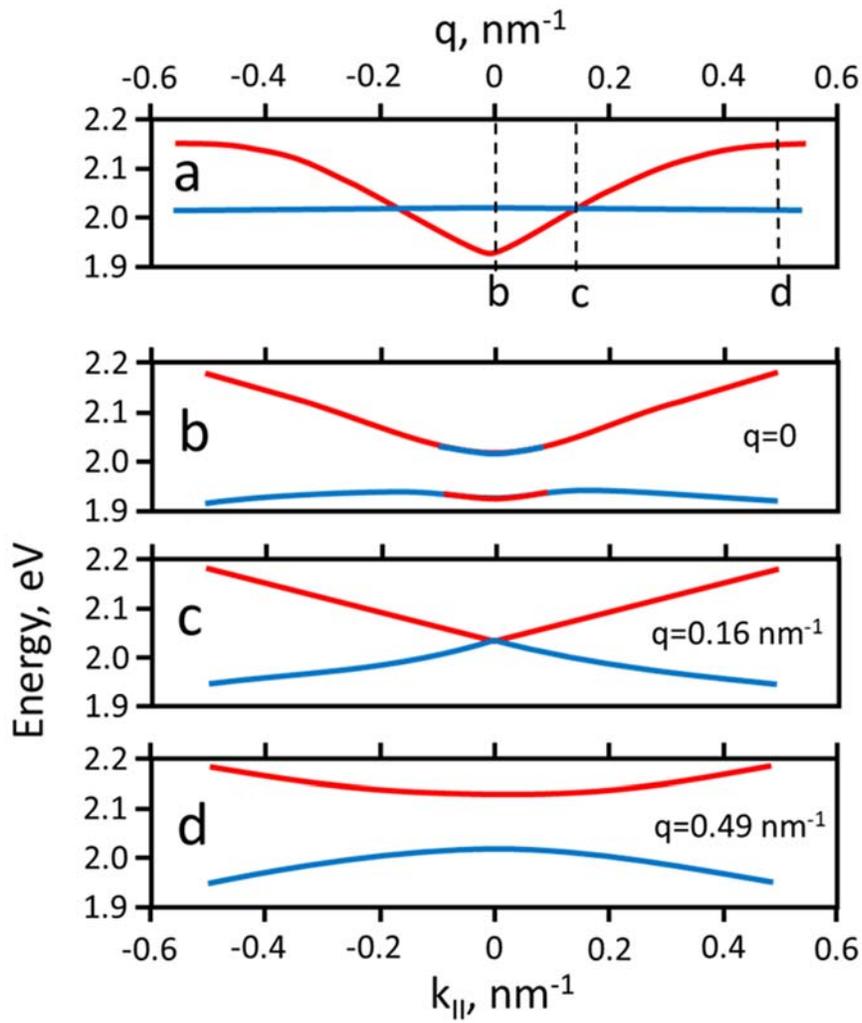